\documentclass[12pt,preprint]{aastex}

\newcommand{\Msun}{{\mbox{$\rm\, M_{\odot}$}}}
\newcommand{\kTe}{$kT_{\rm e}$} 
\newcommand{\rX}{reduced-$\chi^2$} 

\shorttitle{The plasma structure of the Cygnus Loop}
\shortauthors{Tsunemi et al.}

\begin{document}

%% LaTeX will automatically break titles if they run longer than
%% one line. However, you may use \\ to force a line break if
%% you desire.

\title{The Plasma Structure of the Cygnus Loop from the Northeastern
Rim to the Southwestern Rim} 

%% Use \author, \affil, and the \and command to format
%% author and affiliation information.
%% Note that \email has replaced the old \authoremail command
%% from AASTeX v4.0. You can use \email to mark an email address
%% anywhere in the paper, not just in the front matter.
%% As in the title, use \\ to force line breaks.

\author{Hiroshi Tsunemi\altaffilmark{1}, Satoru Katsuda\altaffilmark{1},
Norbert Nemes\altaffilmark{1}, and Eric D. Miller\altaffilmark{2}}
%\affil{Department of Earth and Space Science, Graduate
%  School of Science, Osaka University, 1-1 Machikaneyama, Toyonaka,
%  Osaka 560-0043, Japan}
\email{tsunemi@ess.sci.osaka-u.ac.jp, katsuda@ess.sci.osaka-u.ac.jp,
nnemes@ess.sci.osaka-u.ac.jp, milleric@space.mit.edu}   

%% Notice that each of these authors has alternate affiliations, which
%% are identified by the \altaffilmark after each name.  Specify alternate
%% affiliation information with \altaffiltext, with one command per each
%% affiliation.

\altaffiltext{1}{Department of Earth and Space Science, Graduate
  School of Science, Osaka University, 1-1 Machikaneyama, Toyonaka,
  Osaka 560-0043, Japan}
\altaffiltext{2}{Kavli Institute for Astrophysics and Space Research,
  Massachusetts Institute of Technology, Cambridge, MA 02139, U.S.A.}

%% Mark off your abstract in the ``abstract'' environment. In the manuscript
%% style, abstract will output a Received/Accepted line after the
%% title and affiliation information. No date will appear since the author
%% does not have this information. The dates will be filled in by the
%% editorial office after submission.

\begin{abstract}
The Cygnus Loop was observed from the northeast to the southwest with
 XMM-Newton.  We divided the observed region into two parts, the north
 path and the south path, and studied the X-ray spectra along two paths.
 The spectra can be well fitted either by a one-component
 non-equilibrium ionization (NEI) model or by a two-component NEI model.
 The rim regions can be well fitted by a one-component model with
 relatively low \kTe~whose metal abundances are sub-solar (0.1--0.2).
 The major part of the paths requires a two-component model.  Due to
 projection effects, we concluded that the low \kTe~($\sim$0.2\,keV)
 component surrounds the high \kTe~($\sim$0.6\,keV) component, with the
 latter having relatively high metal abundances ($\sim$5 times solar).
 Since the Cygnus Loop is thought to originate in a cavity explosion,
 the low \kTe~component originates from the cavity wall while the high
 \kTe~component originates from the ejecta.  

The flux of the cavity wall component shows a large variation along our
 path. We found it to be very thin in the south-west region, suggesting a
 blowout along our line of sight.  The metal distribution inside the
 ejecta shows non-uniformity, depending on the element.  O, Ne and Mg are
 relatively more abundant in the outer region while Si, S and Fe are
 concentrated in the inner region, with all metals showing strong
 asymmetry. This observational evidence implies an asymmetric
 explosion of the progenitor star. The abundance of the ejecta also
 indicates the progenitor star to be about 15\,\Msun.  
\end{abstract}

%% Keywords should appear after the \end{abstract} command. The uncommented
%% example has been keyed in ApJ style. See the instructions to authors
%% for the journal to which you are submitting your paper to determine
%% what keyword punctuation is appropriate.


\section{Introduction}

A supernova remnant (SNR) reflects the abundance of the progenitor star when
the remnant is young and that of the interstellar matter (ISM) when it
becomes old.  In this way, we can study the evolution of the ejecta and
the ISM.  The Cygnus Loop is a proto-typical middle-aged shell-like SNR.
The angular diameter is about 2$^{\circ}$.4 and it is very close to us
(540\,pc; Blair et al.\ 2005), implying a diameter of $\sim$23\,pc.
The estimated age is about 10000\,years, less than half that based on
the previous distance estimate of 770\,pc \cite{minkowski58}.  

Since the Cygnus Loop is an evolved SNR, the bright shell mainly
consists of a shock-heated surrounding material.  Its supernova (SN)
explosion is generally considered to have occurred in a preexisting
cavity \cite{mccray79}.  Levenson et al. (1997) found that the Cygnus
Loop was a result of a cavity explosion that was created by a star no
later than B0.  It is almost circular in shape with a break-out in the
south where the hot plasma extends out of the circular shape.  Miyata et
al. (1994) observed the northeast (NE) shell of the Loop with ASCA and
revealed the metal deficiency there \cite{miyata94}.  Since Dopita et
al. (1977) reported the metal deficiency of the ISM around the Cygnus
Loop, they concluded that the plasma in the NE-shell is
dominated by the ISM.  Due to the constraints of the detector efficiency,
they assumed that the relative abundances of C, N and O are equal to
those of the solar value \cite{anders89}.  More recently, Miyata et
al. (2007) used the Suzaku satellite 
\cite{mitsuda07} to observe one pointing position in
the NE rim.  They detected emission lines from C and N and determined
the relative abundances \cite{miyata07}.  They concluded that the
relative abundances of C, N and O are consistent with those of the solar
values whereas the absolute abundances show depletion from the solar
values \cite{anders89}.  Katsuda et al. (2007)
observed four pointings in the NE rim and detected a region where the
relative abundances of C and N are a few times higher than that of O.  

Hatsukade \& Tsunemi (1990) detected a hot plasma inside the Cygnus Loop
that is not expected in the simple Sedov model \cite{hatsukade90}.  They
reported that the hot plasma was confined inside the Loop.  Miyata et
al. (1998) detected strong emission lines from Si, S and Fe-L from
inside the Loop \cite{miyata98}.  They found that the metal abundance is
at least several times higher than that of the solar value
\cite{anders89}, indicating 
that a few tens of higher than that of the shell region.  They
concluded that the metal rich plasma was a fossil of the SN explosion.
The abundance ratio of Si, S and Fe indicated the progenitor star mass
to be 25\Msun.  Miyata \& Tsunemi (1999) measured the radial profile
inside the Loop and found a discontinuity around 0.9\,R$_\mathrm{s}$
where R$_\mathrm{s}$ is the shock radius.  They measured
the metallicity inside the hot 
cavity and estimated the progenitor mass to be 15\Msun.  Levenson et
al. (1998) estimated the size of the cavity and the progenitor mass to
be 15\Msun.  Therefore, the progenitor star of the Cygnus Loop is a
massive star in which the triple-$\alpha$ reaction should have dominated
rather than the CNO cycle.  If the surrounding material of the Cygnus Loop
is contaminated by the stellar activity of the progenitor star, it may
explain the C abundance inferred for this region with Suzaku
\cite{katsuda07}.   

In order to study the plasma condition inside the Cygnus Loop, we
observed it from the NE rim to the south-west (SW) rim with the
XMM-Newton satellite.  We report here the result covering a full
diameter by seven pointings.   

\section{Observations}

We performed seven pointing observations of the Cygnus Loop so that we
could cover the full diameter from the NE rim to the SW rim (from Pos-1
to Pos-7) during the AO-1 phase.  We concentrate on the data obtained
with the EPIC MOS and PN cameras.  All the data were taken by using
medium filters and the prime full window mode.  Fortunately, all the
data other than Pos-4 suffered very little from background flares.
Obs IDs, the observation date, the nominal point, and the effective
exposure times after rejecting the high-background periods are
summarized in table~\ref{obs}.     

All the raw data were processed with version 6.5.0 of the XMM Science
Analysis Software (XMMSAS).  We selected X-ray events corresponding to
patterns 0--12 and 0 for MOS and PN, respectively.  We further cleaned
the data by removing all the events in bad columns listed in the
literature~\cite{kirsch06}.  After filtering the data, they were
vignetting-corrected using the XMMSAS task {\tt evigweight}.  For the 
background subtraction, we employed the data set accumulated from blank
sky observations prepared by \cite{read03}.  After adjusting its
normalization to the source data by using the energy range between
5\,keV and 12\,keV, where the emission is free from the
contamination~\cite{fujita04, sato05}, we subtracted the background data
set from the source.

\section{Spatially Resolved Spectral Analysis}

\subsection{Band image}

Figure~\ref{image} displays an exposure-corrected ROSAT HRI image of the
entire Cygnus Loop (black and white) overlaid with the XMM-Newton color
images of the merged MOS1/2, PN data from all the XMM-Newton
observations.  In this figure, we allocated color codes as red
(0.3--0.52\,keV), green (0.52--1.07\,keV) and blue (1.07--3\,keV).  We
see that the outer regions are reddish rather than 
bluish while the central region is in bluish.   

The NE rim is the brightest in our field of view (FOV), showing a bright
filament at $45^{\circ}$ to the radial direction corresponding to
NGC6992.  The SW rim is also bright in our FOV where there is a V-shape
structure \cite{aschenbach99}.  In the center of the Loop, an X-ray
bright filament runs through Pos-4 and Pos-5 forming a circular
structure.  In the ROSAT image, we can see it and find that it forms a
large circle within the Cygnus Loop.  In this way, there are many fine
bright filaments in intensity.  However, we find that there is a clear
intensity variation along our scan path: dim in the center and bright in
the rim.  

\begin{figure}
\includegraphics[angle=0,scale=1]{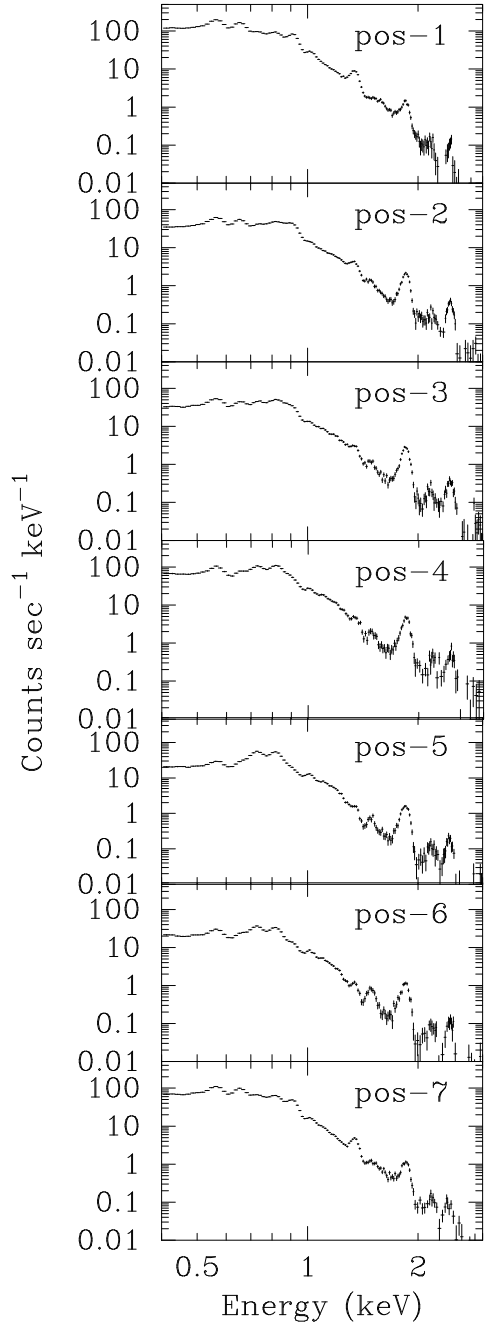}
\caption{MOS1 spectra for the seven pointings; each is the sum of the entire FOV.}
\label{all_spectra}
\end{figure}

\begin{figure}
\includegraphics[angle=0,scale=.5]{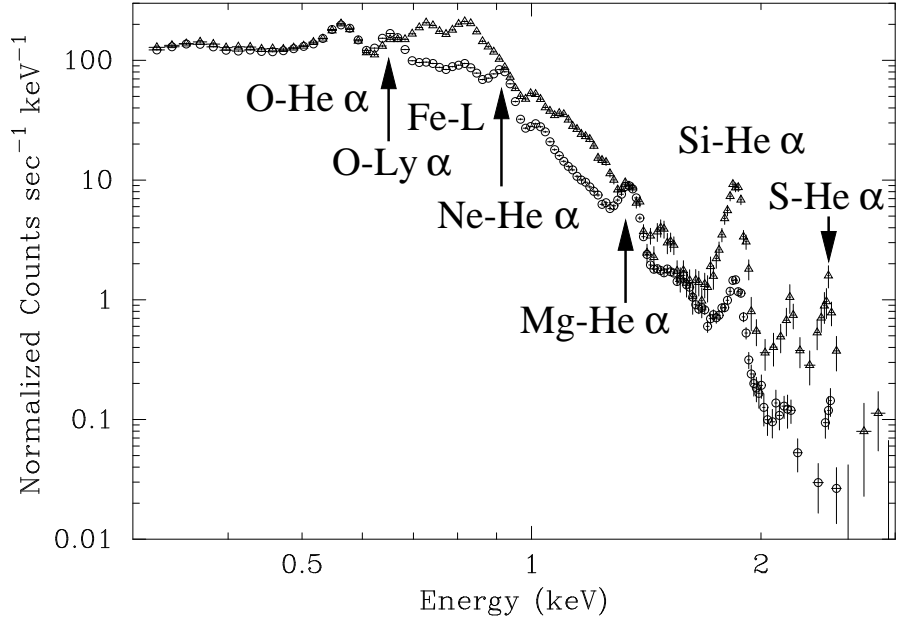}
\caption{Comparison of spectra between Pos-1 (circles) and Pos-4
 (triangles).  They are equalized in intensity at O-He$\alpha$.
}
\label{spec}
\end{figure}

Figure~\ref{all_spectra} shows spectra for seven pointings; each is the
sum of the entire FOV.  The NE rim (Pos-1) and the SW rim (Pos-7) show
strong emission lines below 1\,keV including O, Fe-L and Ne, while the
center (Pos-4) shows strong emission lines from Si and S.  We can see
that the equivalent width of Si and S emission lines are bigger in the
center and gradually decrease toward the rim.  We show the comparison of
spectra between Pos-1 and Pos-4 in figure~\ref{spec}.  Prominent
emission lines are O-He$\alpha$, O-Ly$\alpha$, Fe-L complex,
Ne-He$\alpha$, Mg-He$\alpha$, Si-He$\alpha$, and S-He$\alpha$.  We see
that the emission line shapes for O are quite similar to each other
while there is a big difference at higher energy band.  Since the
spectrum in the NE rim can be well represented by a single temperature
plasma model~\cite{miyata94}, we need an extra component in the center.  

\subsection{Radial profile}

\begin{figure}
\includegraphics[angle=0,scale=.45]{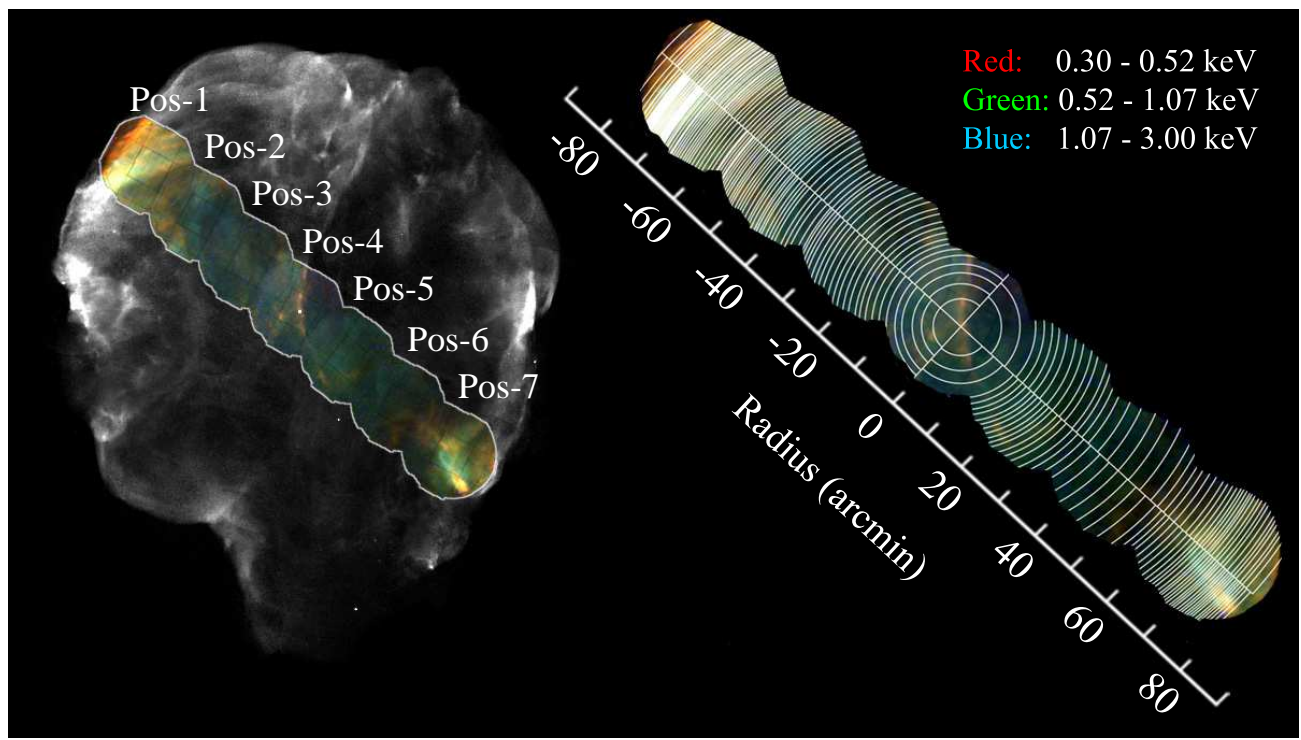}
\caption{Left: Exposure-corrected ROSAT HRI image of the entire Cygnus
 Loop (black and white) overlaid with the XMM-Newton color images of the
 merged MOS1/2, PN data from all the XMM-Newton observations.  
 Right: Spectral extraction regions overlaid on the XMM-Newton 
 three-color image shown in left figure.} 
\label{image}
\end{figure}

Although there are many fine structures, no matter how finely we
divide our FOV, each region would contain different plasma conditions
due to the integration of the emission along the line of sight.
Therefore, we concentrate on large scale structure along the scan path.
First of all, we divided our FOV into two parts along the diameter: the
north path and the south path.  Then we divided them into many small
annular sectors whose center is located at (20$^h$51$^m$34$^s$.7,
31$^\circ$00$^\prime$00$^{\prime\prime}$), i.e., the center of the
nominal position of Pos-4.  There are 141 and 172 annular sectors for
the north path and the south path, respectively.  These small annular
sectors, shown in figure~\ref{image}, are divided such that each has at
least 60,000\,photons ($\sim$20,000 for MOS1/2 and $\sim$40,000 for PN)
to equalize the statistics.  We extracted the spectrum from each sector
using the data set accumulated from blank sky observations as sky
background.  We have confirmed that the emission above 3\,keV is
statistically zero.   In this way, we obtained 313 spectra.  These
sectors can be identified by the angular distance, ``R'', from the
center (east is negative and west is positive as shown in
figure~\ref{image}).    

The width of the sector depends on R.  The sector widths range from
3$^\prime$.8 to 0$^\prime$.2 in the north path and from
3$^\prime$.0 to 0$^\prime$.2 in the south path.  The widest sectors are
in Pos-4 due to its short exposure because of the background flare.  The
narrowest sectors are in the NE rim where the surface brightness is the
highest.   

\subsection{Single temperature NEI model}

\begin{figure}
\includegraphics[angle=0,scale=.5]{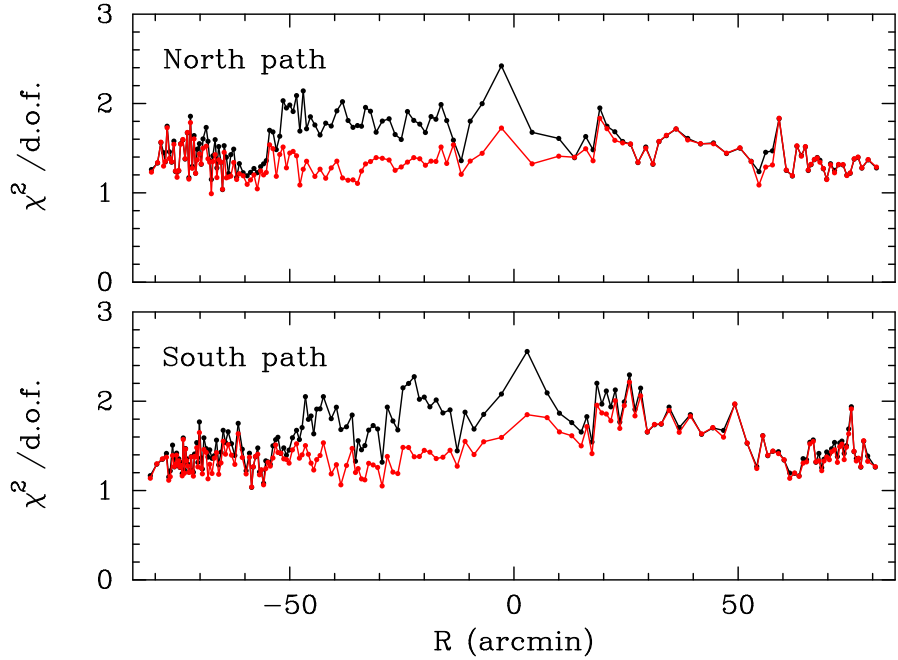}
\caption{Distribution of the \rX\, as a function of R along the
 north path (upper panel) and the south path (lower panel).  The
 single-component model is shown in black and the two-component model is
 shown in red.}  
\label{chi2_dist}
\end{figure}

We fitted the spectrum for each sector with an absorbed non-equilibrium
ionization (NEI) model with a single $kT_{\rm e}$, using the Wabs
\cite{morrison83} and VNEI model (NEI version 2.0) \cite{borkowski01} in
XSPEC v\,12.3.1 \cite{arnaud1996}).  We fixed the column density,
$N_\mathrm{H}$ to be $4.0\times10^{20} \mathrm{cm}^{-2}$ (e.g., \cite{inoue80},
\cite{kahn80}).  Free parameters were \kTe; the ionization time scale, $\tau$
(a product of the electron density and the elapsed time after the shock
heating); the emission measure (hereafter EM; EM = $\int n_{\rm e}n_{\rm
H} d\ell$, where $n_{\rm H}$ and $n_{\rm e}$ are the number densities of
hydrogens and electrons, $d\ell$ is the plasma depth); and abundances of C,
N, O, Ne, Mg, Si, S, Fe, and Ni.  We set abundances of C and N equal to
that of O, that of Ni equal to Fe, and other elements fixed to the
solar values~\cite{anders89}.  In the fitting process, we set 20 as the
minimum counts in each spectral bin to perform the $\chi^2$ test.  We
determined the value of the minimum counts such that it did not affect
the fitting results.    Figure~\ref{chi2_dist} shows the distribution of
the \rX\, in black as a function of R along both the north path and
the south path.  We found that values of \rX~ for all the sectors
are between 1.0 and 2.0.  If we took into account a systematic
error~\cite{nevalainen03, kirsch04} of 5\,\%, the \rX~was around 1.5 or
less.  

In general, values of the \rX\, are a little higher in the central part
of the Cygnus Loop.  Miyata et al. (1994) observed the NE rim with ASCA
and found that the spectra were well represented with a one-temperature
VNEI model with a temperature gradient towards the inside.  The Suzaku
observation in the NE rim~\cite{miyata07} reveals that the X-ray
spectrum can be represented by a two-temperature model: one component is
0.2--0.35\,keV and the other is 0.09--0.15\,keV.  In our fitting, the value of
\kTe~obtained is 0.2--0.25\,keV.  Therefore, we detected a hot component
that Suzaku detected.  There may be an additional component with low
temperature that seems difficult to detect with
XMM-Newton due to the relatively lower sensitivity below 0.5\,keV
compared to Suzaku. 

\begin{figure}
\includegraphics[angle=0,scale=.5]{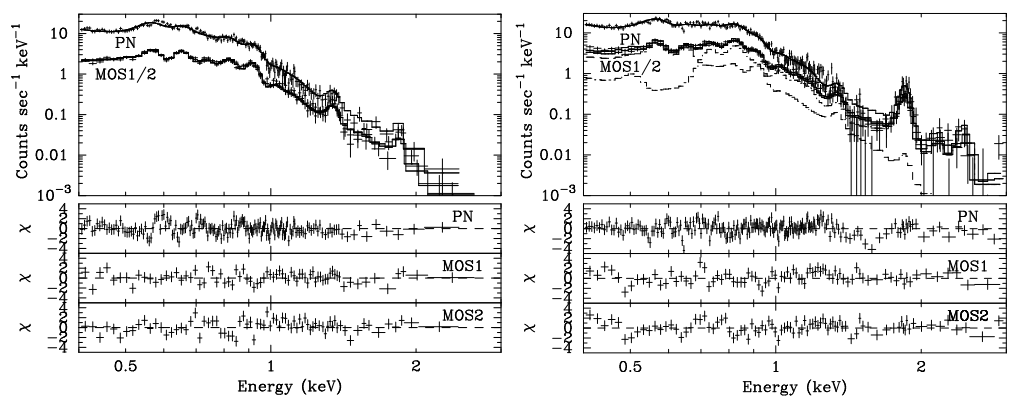}
\caption{Left: An example spectrum that comes from the sector at
 R=$-74.25^\prime$.  The best-fit curves are shown with solid lines and
 the lower panels show the residuals.  Right: Same as left but for the
 sector at R=$+10^\prime$.  Both the ejecta and cavity components are
 shown only for MOS1 spectrum as dashed lines.} 
\label{ex_spec}
\end{figure}

The ASCA observation \cite{miyata94} also shows that the metal abundance
in the NE rim is deficient.  The authors concluded that the plasma in
the NE-rim 
consists of the interstellar matter (ISM) rather than the ejecta.  This
is confirmed with the Suzaku observation \cite{miyata07} that indicates the
abundances of C, N and O to be $\sim$0.1, 0.05 and 0.1 solar , respectively.
We also obtained the metal deficiency in the NE rim data; the best-fit
results are given in figure~\ref{ex_spec} (left) and in table~\ref{param1}.
Leahy (2004) measured the X-ray spectrum of the southwest region of the
Cygnus Loop and reported that the oxygen abundance there is about 0.22
solar \cite{leahy04}.  Therefore, the X-ray measurements of the Cygnus
Loop show that the metal abundances are depleted. 

Cartledge et al.\ (2004) measured the interstellar oxygen along 36 sight
lines and confirmed the homogeneity of the O/H ratio within 800\,pc of
the Sun.  We found that they measured it in the direction about
$5^\circ$ away from the Cygnus Loop.  The oxygen abundance they measured is
about 0.4 times the solar value \cite{anders89}.  Wilms et
al.\ (2000) employed 0.6 of the total interstellar abundances for the
gas-phase ISM oxygen abundance, and suggest that this depletion may be
due to grains.  Although the ISM near the Cygnus Loop may be
depleted, the abundances are still much higher than what we obtained at
the rim of 
the Cygnus Loop. It is difficult to explain such a low abundance of
oxygen in material originating from the ISM.  Therefore, the
origin of the low metal abundance is open to the question.  Since the
Cygnus Loop is thought to have exploded in a pre-existing cavity, we
can say that the cavity material shows low metal abundance.  The
abundance difference between our data and those from Suzaku may be due
to the difference in the detection efficiency at low energy.  Taking
into account the projection effect, the plasma of the rim regions
consists only of the cavity material while that of the inner regions
consists both of the cavity material and of an extra component filling
the interior of the Loop.   

\subsection{Two-temperature NEI model}

To further constrain the plasma conditions, we applied a two-component
NEI model with different temperatures. In this model, we added an extra
component to the single 
temperature model.  The extra component is also an absorbed VNEI model
with \kTe, $\tau$, and EM as free parameters.  The metal
abundances of the extra component are fixed to those determined at the
NE rim so that the extra component represents the cavity material.
Figure~\ref{chi2_dist} shows \rX~values in red along the path.  Applying
the $F$-test with a significance level of 99\% to determine whether or
not an extra component is needed, we found that most of the spectra
required a two-component model, particularly in the central part of the
Loop.  Sectors that do not require two-component model are mainly
clustered in R$<$-65$^\prime$,  +25$^\prime<$R$<+40^\prime$, and
+60$^\prime<$R.  Therefore, we considered that the outer sectors
($\mid$R$\mid >70^\prime$) can be safely represented by a one-component
model while other sectors can be represented by a two-component model.  In
this way, we performed the analysis by applying a two-component VNEI
model with different temperatures.  We assumed that the low temperature
component comes from the surrounding region of the Cygnus Loop and the
high temperature component occupies the interior of the Loop.    

We found that the values of \rX~are 1.0--1.8 even employing a
two-component model.  This is partly due to the systematic
errors. Looking at the image in detail, there are fine structures within
the sector.  Furthermore, the spectrum from each sector is an
integration along the line of sight.  Since we only employ two VNEI
plasma models, the values of \rX~ are mainly due to the simplicity of the
plasma model employed here.  Therefore, we think that the plasma
parameters obtained will represent typical values in each sector.  

Figure~\ref{ex_spec} (right) and table~\ref{param2} shows an example
result that comes from the sector at R=$+10^\prime$.  Fixed parameters
in the low \kTe~component come from the fitting result at the NE rim
obtained by Suzaku observations \cite{Uchida06}.  Metal abundances for
the high \kTe~component show higher values by an order of magnitude than
those of the low \kTe~component, surely confirming that the high
\kTe~component is dominated by fossil ejecta. 

\begin{figure}
\includegraphics[angle=0,scale=.5]{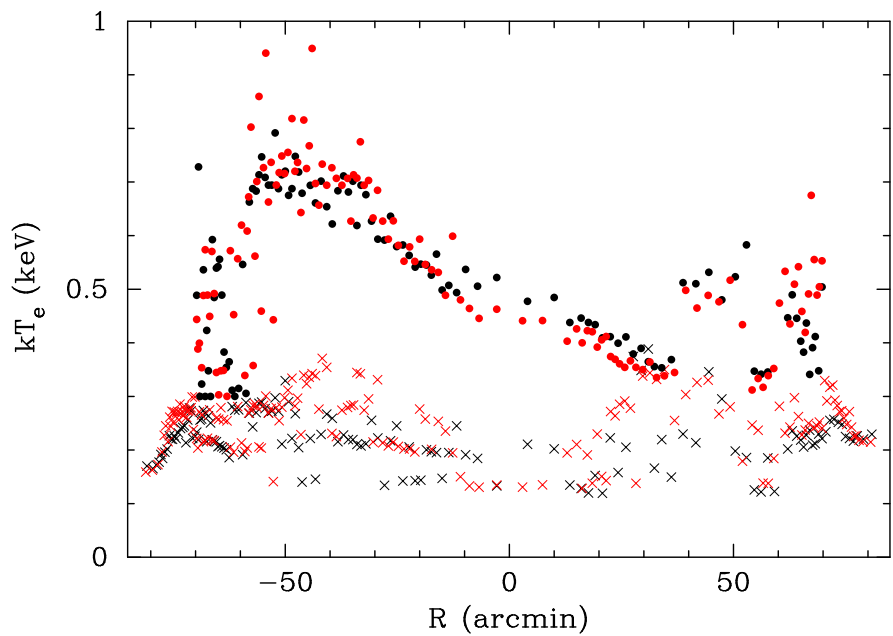}
\caption{Temperature distributions of the two components as a function of
 position.  
 Filled circles show the ejecta component, while crosses show the cavity
 component. Black show the north path and red shows the south path.
 Typical errors are $\pm$5\% for both components.
}
\label{kT_dist}
\end{figure}

\begin{figure}
\includegraphics[angle=0,scale=.5]{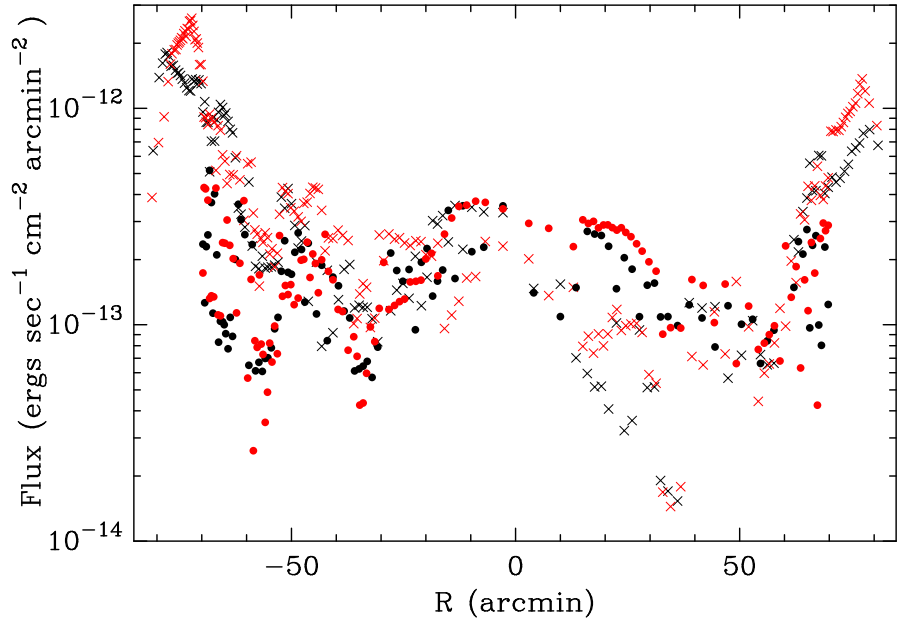}
\caption{Flux distributions of the two components as a function of position.
 The marks in this figure are the same as those in
 figure~\ref{kT_dist}.} 
\label{flux_dist}
\end{figure}

Figure~\ref{kT_dist} shows temperatures as a function of position.  The
low \kTe~component is in the temperature range of 0.12--0.34\,keV while
the high \kTe~component is above 0.35\,keV.  There is a clear temperature
difference where a two-component model is required rather than a single
temperature model.  The low \kTe~component represents the cavity
material surrounding the Cygnus Loop while the high \kTe~component
represents the fossil ejecta inside the Loop.
Figure~\ref{flux_dist} shows the fluxes for the two components as a function
of position.  The low \kTe~component shows clear rim brightening.  The
east part is stronger than the west part, showing asymmetry of the Loop.
On the other hand, the high \kTe~component has a relatively flat radial
dependence.  From the center to the SW, we see that the flux of the high
\kTe~component is stronger than that of the low \kTe~component.  

\subsubsection{Distribution of the cavity material}

As shown in figure~\ref{kT_dist}, the low \kTe~component shows
relatively constant temperature with radius.  The distribution of the
flux shown in figure~\ref{flux_dist} shows peaks in the rim and
relatively low values inside the Loop.  There are some differences
between the north path and the south path.  The biggest one is a clear
difference 
in peak position in the NE rim that is due to the bright filament at
$45^{\circ}$ to the radial direction, as seen in
figure~\ref{image}.  However, these two paths show a globally similar
behavior in flux.  Therefore, we can see that they are quite similar to
each other from a large scale point of view.   

We notice that there are many aspects showing asymmetry and
non-uniformity.  The NE half is stronger in intensity than the SW half.
The flux in the inner part of the Loop shows relatively small values in
the west half, particularly at +25$^\prime<$R$<+40^\prime$.  The NE half
is brighter by a factor of 5--10 than the SW half.  Furthermore,
the SW half shows stronger intensity variation than the NE half.
This suggests that the thickness of the cavity shell is far from
uniform.  The cavity shell in the SW half is much thinner than that in
the NE half.  Since we assumed the metal abundances of the low 
\kTe~component equal to those of the NE rim, we can calculate the EM.
Furthermore, we assumed the ambient density to be 0.7\,cm$^{-3}$ based
on the observation of the NE rim, and we estimate the mass of the low
\kTe~component to be 130\,\Msun. However, we should note that there are
is evidence that the SN explosion which produced the Cygnus
Loop occurred within a preexisting cavity (e.g., Hester et al.\ 1994;
Levenson et al.\ 1998; Levenson et al.\ 1999).  The model predicts that the
original cavity density, $n_c$, is related to the wall density
$n_s$ by $n_c = 5$.  Assuming that $n_0$ equals the ambient density,
$n_s$, we estimate $n_c$ to be 0.14 cm$^{-3}$.  Then, we calculate the total
mass in the preexisting cavity to be $\sim$25 M$_\odot$.

\subsubsection{Ejecta distribution}

\begin{figure}
\includegraphics[angle=0,scale=.8]{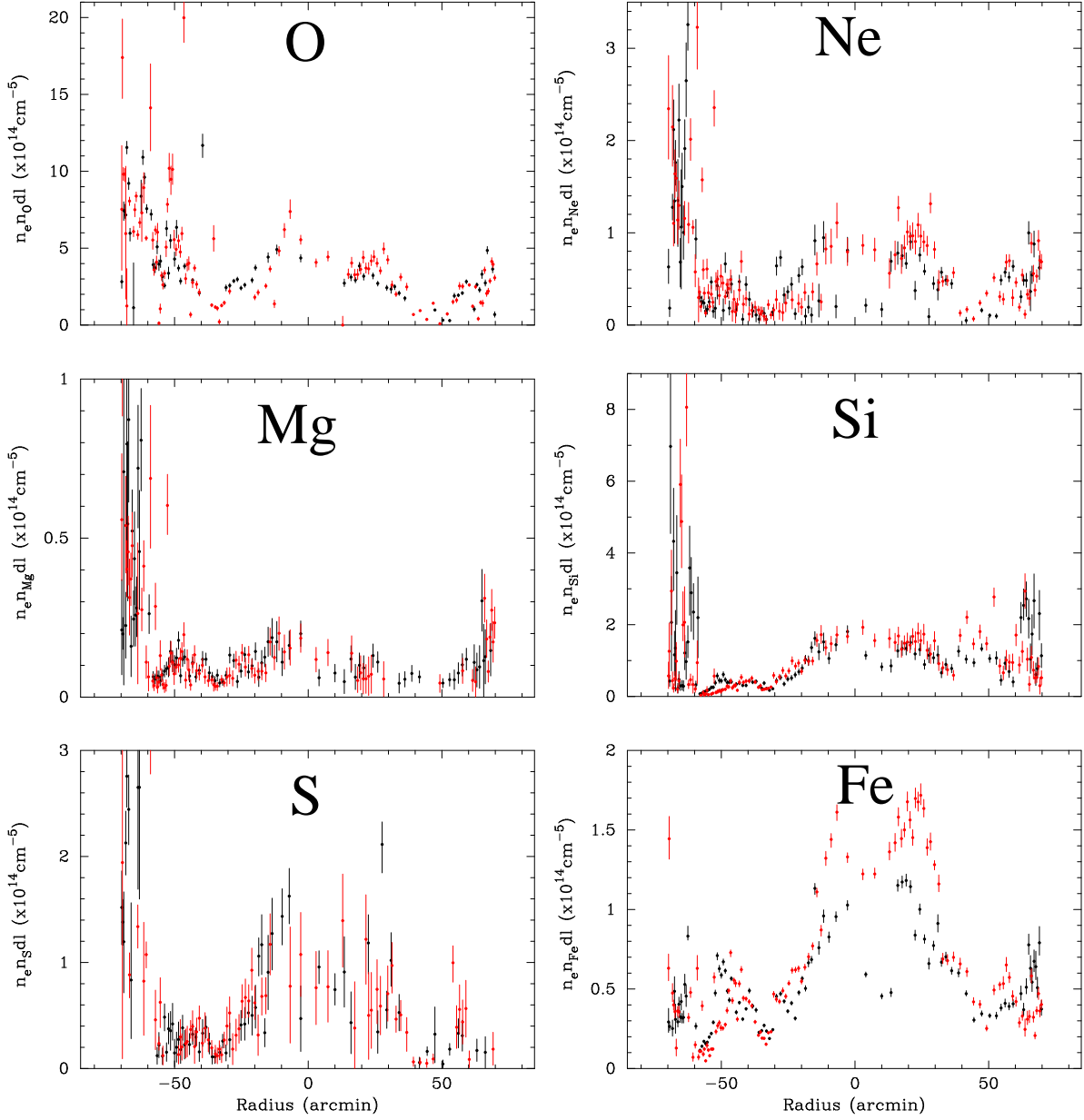}
\caption{Distributions of EM for various metals (O(=C=N), Ne, Mg, Si, S,
 and Fe(=Ni)) in the ejecta.  Black indicates the north path
 and red indicates the south path.  Data points showing only upper limits are excluded
 in these figures.} 
\label{metal_dist}
\end{figure}

The flux distribution from the ejecta along the path is shown with filled
circles in figure~\ref{flux_dist}.  It has a relatively flat structure
with two troughs around R=$-35^\prime$ and R=+50$^\prime$.  Since we
left the metal abundances as free parameters, we obtained distributions
of EM of various metals (C=N=O, Fe=Ni, Ne, Mg, Si and S) in the ejecta.
These are shown in figure~\ref{metal_dist}, where black crosses trace
the north path and red crosses trace along the south path.  If we assume
uniform plasma conditions along the line of sight, the EM represents the
mass of the metal.  Most elements show similar structure between the
north path and the south path while there is a big discrepancy in Fe/Ni
distribution at $-10^\prime <$ R $< +30^\prime$.  In this region, the
south path shows two times more abundant Fe/Ni than the north path.  A
similar discrepancy is seen in O ($-30^\prime <$ R $< -10^\prime$) and
in Ne (at $-10^\prime <$ R $< +10^\prime$).  Therefore, the distribution
of metal abundance shows a north-south asymmetry along the path.  

The distributions of O and Ne show a central bump and increase at the
outer sectors.  However, those of Mg, Si, S and Fe only show a central
bump.  The increase of O and Ne at the outer sectors indicates that the
outer parts of the ejecta mainly consists of O and Ne and they may be
well mixed.  Similarly heavy elements, Mg, Si, S and Fe/Ni forming
central bumps may show that they are well mixed.  Therefore, a
significant convection has occurred in the central bumps while an
``onion-skin'' structure remains in the outer sectors. 

\section{Discussion}
  
The Cygnus Loop appears to be almost circular with a blowout in the
south.  The ROSAT image indicates no clear shell in this blowout
region.  Levenson et al. (1997) revealed that there is a thin shell left
at the edge of the blowout region.  Therefore, there is a small amount
of cavity material in this region that surrounds the ejecta.  This also
indicates the non-uniformity of the cavity wall.  If the cavity wall is
thin, the ejecta can produce a blowout structure. 

Looking at the component of the cavity material along our path shown in
figure~\ref{flux_dist}, the flux is very weak at
+15$^\prime<$R$<+40^\prime$.  This indicates that the cavity wall is
very thin in this region.  When we calculate the flux ratio between the
ejecta plasma and the cavity material, we find that the ratio becomes
high (larger than 4) at +15$^\prime<$R$<+35^\prime$ in the north path
and +30$^\prime<$R$<+35^\prime$ in the south path.  Therefore, we guess that
the thin shell region is larger in the north path than in the south
path.  This also shows the asymmetry between the north and the south as
well as that between the east and the west.  If the thin shell region
corresponds to a blowout similar to that in the south blowout, this
region must have a blowout structure along the line of sight either in
the near side or far side or both.  This structure roughly corresponds
to Pos-5 and will extend further in the northwest direction.  Looking at
the ROSAT image in figure~\ref{image}, we see a circular region with low
intensity.  It is centered at (20$^h$49$^m$11$^s$,
31$^\circ$05$^\prime$20$^{\prime\prime}$) with a radius of 30$^\prime$.
We guess that this circular region corresponds to a possible blowout in
the direction of our line of sight.  CCD observation just north of
our path will answer this hypothesis.   

We obtained EMs of O, Ne, Mg, Si, S, and Fe for the ejecta along the
north path and the south path.  Multiplying the EMs by the area of each
sector, we obtained emission integral (hereafter EI, EI$=\int
n_\mathrm{e}n_\mathrm{H} dV$, $dV$ is the X-ray-emitting volume) along
the path.  Since we only observed the limited area of the Cygnus Loop
from the NE rim to the SW rim, we have to estimate the EIs for the
entire remnant in order to obtain the relative abundances as well as the
total mass of the ejecta.  Therefore, we divide our observation region
into four parts: left-north part, right-north part, left-south part, and
right-south part.  We assume that each part represents the average EIs
of the corresponding quadrant of the Loop.   
In this way, we can calculate the total EIs for O, Ne, Mg, Si, S, and Fe
that are described in table~\ref{emission_integral}.  The south
quadrant, corresponding to the right-south path, contains the largest
mass fraction of 31\%, while the other quadrants contain 23\% each.  Then, we
calculate the relative abundances of Ne, Mg, Si, S, and Fe to O in the
entire ejecta.  Since we cannot measure the abundance of light elements
like He, it is quite difficult to estimate the absolute abundances.
However, the relative abundance to O is robust.  

\begin{figure}
\includegraphics[angle=0,scale=.5]{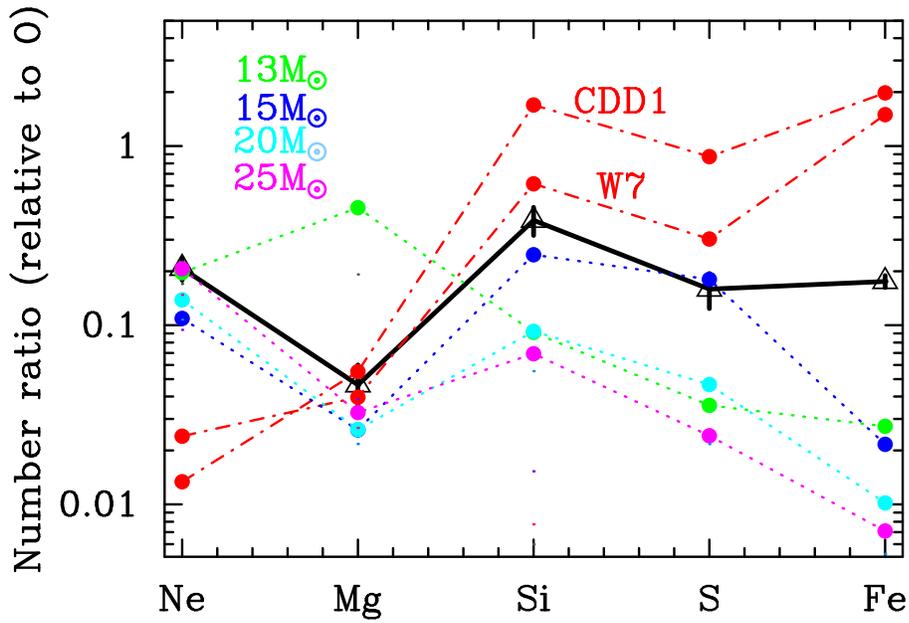}
\caption{Number ratios of Ne, Mg, Si, S, and Fe relative to O of the
  high-$kT_{\rm e}$ component estimated for the entire Loop (black solid
 line).  Dash-dotted red lines represent CDD1 and W7
 model of Type-Ia \cite{iwamoto99}.  Dotted green, blue, light blue, and
 magenta lines represent core-collapse models whose progenitor masses
 are 13, 15, 20, and 25 \Msun, respectively \cite{woosley95}. } 
\label{rel_abund}
\end{figure}

Since the Cygnus Loop is believed to be a result from a core-collapse
SN, we compared our data with core-collapse SN models.  There are many
theoretical results from various authors (e.g.,
Woosley \& Weaver 1995; Thielemann et al.\ 1996; Rauscher et al.\ 2002;
Tominaga et al.\ 2007).  We also employed a SN Type Ia model
\cite{iwamoto99} for comparison.  We calculated the relative
abundance for various elements to O and compared them with models.
Figure~\ref{rel_abund} shows comparisons between the model
calculations and our results where we picked up Woosley's model with
one solar abundance \cite{anders89} for the core-collapse case
\cite{woosley95}.  The Type Ia model yields more Si, S and Fe than our
results, but less Ne. 
Models with massive stars produce better fits to our results than the Type
Ia model.  Among them, we found that the model with 15\,\Msun~showed
good fits to our results.  They fit within a factor of two with an
exception of Fe.  We also noticed that the model with one solar
abundance looked better fit than that with depleted abundance.
Therefore, we can conclude that the Cygnus Loop originated from
an approximately 15\,\Msun~star with one solar abundance.  

Assuming that the ejecta density is uniform along the line of sight, we
estimate the total mass of the fossil ejecta to be 21\,\Msun.  In
this calculation, we assumed that the electron density is equal to that
of hydrogen and that the plasma filling factor is unity, although the
fossil ejecta might be deficient in hydrogen.  If it is the case, the
total mass of the fossil ejecta reduces to $\sim$12,\,\Msun~whereas
the relative abundances are not affected.  The most suitable
nucleosynthetic model predicts that the total mass ejected is about
6\,\Msun~without H.  Therefore, there might be a significant amount of
contamination from the swept-up matter into the high-\kTe~component,
which we consider the ejecta.  Otherwise, the assumption that the
density of the ejecta is uniform might be incorrect since 
rim-brightening for the EMs of O, Ne, Mg, and Fe is clearly seen in
Fig.~\ref{metal_dist}. Non-uniformity reduces the filling factor and also the
mass of the high-\kTe~component.   

There is observational evidence of the asymmetry of supernova
explosions both for massive stars~\cite{leonard06} and for Type
Ia~\cite{motohara06}.  We found that the ejecta plasma shows asymmetric
structure between NE half and SW half.  Ne and Fe are evenly divided
while two thirds of O and Mg are in the NE half.  On the contrary, two
third of Si and S are in the SW half.  We calculated the ejecta mass for
each quadrant and found that the south quadrant contains the largest
ejecta mass.  Similar asymmetries are seen in other SNR, such as Puppis A,
which shows asymmetric structure with O-rich, fast-moving knots
(Winkler \& Kirscher 1985; Winkler et al.\ 1988) .  The central compact
object in Puppis~A is on the opposite side of the SNR from the O-rich,
fast-moving knots \cite{petre96}.  If the asymmetry of the ejecta in the
Cygnus Loop is similar to that of Puppis A, we may expect a compact
object to be in the north direction.    

\section{Conclusion}

We have observed the Cygnus Loop along the diameter from the NE rim to
the SW rim employing XMM Newton.  The FOV is divided into two paths: the
north path and the south path.  Then it is divided into many small
annuli so that each annulus contains a similar number of photons to
preserve statistics.  

The spectra from the rim regions can be expressed by a
one-\kTe~component model while those in the inner region require a
two-\kTe~component model.  The low 
\kTe~plasma shows relatively low metal abundance and covers the entire
FOV.  It forms a shell that originates from the preexisting cavity.  The
high \kTe~plasma shows high metal abundance and occupies a large part of
the FOV. The origins of these two components are different: the high
\kTe~plasma with the high metal abundance must come from the ejecta while
low \kTe~plasma with low metal abundance must come from the cavity
material.  We find that the thickness of the shell is very thin in the
south west part where, we guess, the ejecta plasma is blow out in the
direction of our line of sight.  

We estimate the mass of the metals.  Based on the relative metal
abundance, we find that the Cygnus Loop originated from a
15\,\Msun~star.  The distribution of the ejecta is asymmetric,
suggesting an asymmetric explosion.  

\acknowledgments

This work is partly supported by a Grant-in-Aid for Scientific Research
by the Ministry of Education, Culture, Sports, Science and Technology
(16002004).  This study is also carried out as part of the 21st Century
COE Program, \lq{\it Towards a new basic science: depth and
synthesis}\rq.  S. K. is supported by JSPS Research Fellowship for Young
Scientists.

%\end{document}

%\end{document}

\begin{table*}
 \begin{center}
%\tabletypesize{\scriptsize}
 \caption{Summary of the seven observations.}
  \begin{tabular}{lcccc}
\hline\hline
Obs. ID &Camera &Obs. Date& Coordinate (RA, DEC) &Effective Exposure\\
\hline
0082540101 & MOS1 & 2002-11-25 & 20$^h$55$^m$23$^s$.6,
31$^\circ$46$^\prime$17$^{\prime\prime}$.0 & 14.1\,ksec\\
(Pos-1) & MOS2 & & & 14.1\,ksec\\
 & PN & & & 5.6\,ksec\\
\hline
0082540201 & MOS1 & 2002-12-03 & 20$^h$54$^m$07$^s$.4,
31$^\circ$30$^\prime$51.4$^{\prime\prime}$.0 & 14.4\,ksec\\
(Pos-2) & MOS2 & & & 14.4\,ksec\\
 & PN & & & 11.7\,ksec\\
\hline
0082540301 & MOS1 & 2002-12-05 & 20$^h$52$^m$51$^s$.1,
31$^\circ$15$^\prime$25$^{\prime\prime}$.7 & 11.6\,ksec\\
(Pos-3) & MOS2 & & & 11.6\,ksec\\
 & PN & & & 9.1\,ksec\\
\hline
0082540401 & MOS1 & 2002-12-07 & 20$^h$51$^m$34$^s$.7,
31$^\circ$00$^\prime$00$^{\prime\prime}$.0 & 4.9\,ksec\\
(Pos-4) & MOS2 & & & 4.9\,ksec\\
 & PN & & & 3.4\,ksec\\
\hline
0082540501 & MOS1 & 2002-12-09 & 20$^h$50$^m$18$^s$.4,
30$^\circ$44$^\prime$34$^{\prime\prime}$.3 & 12.6\,ksec\\
(Pos-5) & MOS2 & & & 12.6\,ksec\\
 & PN & & & 10.0\,ksec\\
\hline
0082540601 & MOS1 & 2002-12-11 & 20$^h$49$^m$02$^s$.0,
30$^\circ$28$^\prime$16$^{\prime\prime}$.1 & 11.5\,ksec\\
(Pos-6) & MOS2 & & & 11.5\,ksec\\
 & PN & & & 5.9\,ksec\\
\hline
0082540701 & MOS1 & 2002-12-13 & 20$^h$47$^m$45$^s$.8,
30$^\circ$13$^\prime$42$^{\prime\prime}$.9 & 13.5\,ksec\\
(Pos-7) & MOS2 & & & 13.5\,ksec\\
 & PN & & & 7.5\,ksec\\
\hline
\label{obs}
  \end{tabular}
 \end{center}
\end{table*}

\begin{table*}
 \begin{center}
 \caption{Spectral-fit parameters.}
%\tabletypesize{\scriptsize}
  \begin{tabular}{lc}
\hline\hline
Parameter & region$-74.25$\\
\hline
$N_{\rm H} [10^{20}\,{\rm cm}^{-2}$]\dotfill  & 4 (fixed)\\
\hline
$kT_{\rm e}$[keV] \dotfill & 0.23 $\pm$0.01\\
O(=C=N) \dotfill& 0.068 $\pm$0.002\\
Ne\dotfill& 0.17$\pm$0.01 \\
Mg\dotfill& 0.14$\pm$0.03  \\
Si\dotfill& 0.3$\pm$0.1 \\
S\dotfill& 0.6$\pm$0.2 \\
Fe(=Ni)\dotfill& 0.157$\pm$0.006 \\
log$(\tau /\rm cm^{-3}\,sec)$ \dotfill & 11.31 $\pm$0.02\\
EM$^1$[$\times10^{19}$ cm$^{-5}$]\dotfill& 11.0$^{+1.4}_{-0.5}$\\
\hline

$\chi^2$/d.o.f. \dotfill &420/314 \\

\hline
 &  \\%[-10pt]
  \multicolumn{2}{l}{Note. Other elements are fixed to solar values.}\\
\multicolumn{2}{l}{The values of abundances are multiples of solar value.}\\
   \multicolumn{2}{l}{The errors are in the range $\Delta\,\chi^2\,<\,2.7$ on one parameter.}\\
   \multicolumn{2}{l}{$^1$EM denotes emission measure, $\int n_\mathrm{e}n_\mathrm{H} d\ell$.}\\
\label{param1}
  \end{tabular}
 \end{center}
\end{table*}

\begin{table*}
 \begin{center}
 \caption{Spectral-fit parameters.}
%\tabletypesize{\scriptsize}
  \begin{tabular}{lc}
\hline\hline
Parameter & region$+10$\\
\hline
$N_{\rm H} [10^{20}\,{\rm cm}^{-2}$]\dotfill  & 4 (fixed)\\
\hline
\multicolumn{2}{c}{Low temperature component}\\
$kT_{\rm e}$[keV] \dotfill & 0.20 $\pm$0.01\\
C \dotfill& 0.27 (fixed) \\
N \dotfill& 0.10 (fixed) \\
O \dotfill& 0.11 (fixed) \\
Ne\dotfill& 0.21 (fixed)\\
Mg\dotfill& 0.17 (fixed) \\
Si\dotfill& 0.34 (fixed)\\
S\dotfill& 0.17 (fixed)\\
Fe(=Ni)\dotfill& 0.20 (fixed)\\
log$(\tau /\rm cm^{-3}\,sec)$ \dotfill & $12 <$\\
EM$^1$[$\times10^{18}$ cm$^{-5}$]\dotfill& 1.34$^{+0.03}_{--0.04}$\\
\hline

\multicolumn{2}{c}{High temperature component}\\
$kT_{\rm e}$[keV] \dotfill & 0.48 $\pm$0.01\\
O(=C=N) \dotfill& $<$0.01\\
Ne\dotfill& 0.15 $^{+0.06}_{-0.07}$\\
Mg\dotfill& 0.21$\pm$0.08\\
Si\dotfill& 2.5$\pm$0.3\\
S\dotfill& 5$\pm$1\\
Fe(=Ni)\dotfill& 1.03$\pm$0.04\\
log$(\tau /\rm cm^{-3}\,sec)$\dotfill &11.12$\pm$0.05\\
EM$^1$[$\times10^{19}$ cm$^{-5}$]\dotfill& 0.094$^{+0.005}_{-0.004}$\\
\hline
$\chi^2$/d.o.f. \dotfill &531/377 \\

\hline
 &  \\%[-10pt]
  \multicolumn{2}{l}{Note. Other elements are fixed to solar values.}\\
\multicolumn{2}{l}{The values of abundances are multiples of solar value.}\\
   \multicolumn{2}{l}{The errors are in the range $\Delta\,\chi^2\,<\,2.7$ on one parameter.}\\
   \multicolumn{2}{l}{$^1$EM denotes emission measure, $\int n_\mathrm{e}n_\mathrm{H} d\ell$.}\\
\label{param2}
  \end{tabular}
 \end{center}
\end{table*}

\begin{table*}
 \begin{center}
%\tabletypesize{\scriptsize}
 \caption{Calculated emission integral of the Cygnus Loop ejecta.}
  \begin{tabular}{lc}
\hline\hline
Element &  10$^{53}$\,cm$^{-3}$\\
\hline
O   & 7.4$\pm$0.5\\ 
Ne  & 1.5$\pm$0.2 \\
Mg  & 0.34$\pm$0.1\\ 
Si   & 2.9$\pm$0.5 \\
S   & 1.2$\pm$0.3\\ 
Fe  & 1.30$\pm$0.05 \\
\hline
\label{emission_integral}
  \end{tabular}
 \end{center}
\end{table*}

\end{document}